\UseRawInputEncoding
\documentclass[namedreferences]{solarphysics}
\usepackage[hyperref,optionalrh,solaromanenum]{spr-sola-addons} % For Solar Physics 
\usepackage{graphicx}                    % For eps figures, newer & more powerfull
\usepackage{amssymb}                     % useful mathematical symbols                    
\usepackage{color}                       % For color text: \color command
\begin{document}

\begin{article}

\begin{opening}

\title{Analysis of the Solar Flare Index for Solar Cycles 18\,--\,24: Extremely Deep Gnevyshev Gap in the Chromosphere}

%%%%%%%%%%%%%%%%%%%%%%%%%%%%%%%%%%%%%%%%%%%%%%%%%%%
%% Authors Names
%
% \author[addressref={},corref,email={}]{\inits{}\fnm{}\lnm{}\orcid{}}
\author[addressref={aff1},corref,email={jouni.j.takalo@oulu.fi; jojuta@gmail.com}]{\inits{J.J.}\fnm{Jouni}~\lnm{Takalo}}

\institute{$^{1}$ Space Physics and Astronomy Research Unit, University of Oulu,
POB 3000, FIN-90014, Oulu, Finland\\
\email{jouni.j.takalo@oulu.fi}}

%%%%%%%%%%%%%%%%%%%%%%%%%%%%%%%%%%%%%%%%%%%%%%%%%%%
%% Runningheads
%
\runningauthor{Jouni J. Takalo}
\runningtitle{Analysis of SFI}

%%%%%%%%%%%%%%%%%%%%%%%%%%%%%%%%%%%%%%%%%%%%%%%%%%%
%% Affilations 
%% id shold be the same with \author addressref value.
%\address[id={}]{}

%%%%%%%%%%%%%%%%%%%%%%%%%%%%%%%%%%%%%%%%%%%%%%%%%%%
%%% Abstract 

\begin{abstract}

We study the solar flare index (SFI) for the solar cycles 18\,--\,24. We find that SFI has deeper Gnevyshev gap (GG) in its first principal component than other atmospheric parameters. The GG is extremely clear especially in the even cycles.

The GG of the SFI appears about a half year later as a drop in the interplanetary magnetic field near the Earth and in the geomagnetic Ap-index. The instantaneous response of the magnetic field to solar flares, however, shows about two to three days after the eruption as a high, sharp peak in the cross-correlation of the SFI and Ap-index and as a lower peak in SFI vs. IMF B cross-correlation. We confirm these rapid responses using superposed-epoch analysis.

The most active flare cycles during 1944\,--\,2020 are the Cycles 19 and 21. The Cycle 18 has very strong SFI days as much as Cycle 22, but it has least nonzero SFI days in the whole interval. Interestingly Cycle 20 can be compared to the Cycles 23 and 24 in its low flare activity, although it locates between the most active SFI cycles.

\end{abstract}

%%%%%%%%%%%%%%%%%%%%%%%%%%%%%%%%%%%%%%%%%%%%%%%%%%%
%% Keywords
%

\end{opening}
%-------------------------------------------------

%%%%%%%%%%%%%%%%%%%%%%%%%%%%%%%%%%%%%%%%%%%%%%%%%%%
%% Sections
%
% \section{}%\label{s:?} 
\section{Introduction}

Solar flare is a burst of radiation coming from the release of magnetic energy associated with sunspots. Flares are the most energetic phenomena in the Sun. They are seen as bright areas on the Sun and they can last from minutes to hours. The primary ways to monitor flares are in x-rays, energetic particles and optical light.
\cite{Kleczek_1952} quantified the solar flares through a formula Q=Importance $\times$ time to be able approximate the daily flare activity. This quantity, which is defined in more detail in the Data and Methods section, is called the Solar Flare Index (SFI).

The earliest studies of Solar Flare Index (SFI) are based on the recordings of the Astronomical Institute Ond{\v{r}}ejov Observatory of the Czech Academy of Sciences. \cite{Svetska_1956} studied the flares of the Cycles 17 and 18 based on this database. \cite{Knoska_1984} studied the flare activity of the Cycle 20 and \cite{Knoska_1985} the annual distribution of the flares for the interval 1937\,--\,1976 and compared those for the variation of sunspot numbers of the same interval. Most of the studies of SFI have, however, concentrated on spectral studies of the index for the Cycles 20\,--\,24 \citep{Ozguc_1989, Ozguc_2003, Atac_2006, Mendoza_2011, Ozguc_2021}.
\cite{Mendoza_2011} studied also SFI power in the mid-term periodicities (1-2 years) for Cycles 20\,--\,23 and concluded that the flare index is diminished during the low activity cycle 20. The power increases during cycles 21 and 22, but cycle 23 shows again weaker power than Cycles 21 and 22. 

\cite{VelascoHerrera_2022} reconstructed the solar flare index (SFI) to study the solar chromospheric variability from 1937 to 2020. The new SFI database is a composite record of the Astronomical Institute Ond{\v{r}}ejov Observatory of the Czech Academy of Sciences from 1937\,--\, 1976 and the records of the Kandilli Observatory of Istanbul, Turkey from 1977\,--\, 2020. They studied the periodicities of the flare cycles using wavelet transform.
They also found using method called power anomaly, that most active flare cycles were Cycles 17 (incomplete cycle), 19, and 21, while Cycles 20, 22, 23, and 24 were the weakest ones with Cycle 18 was intermediate in flare activity.  Especially, Cycle 20 is much weaker in flare index than in sunspot numbers.

Gnevyshev first noticed that solar cycle has usually twofold maximum, i.e. two maxima with a gap, nowadays called Gnevyshev gap (GG), in between \citep{Gnevyshev_1963, Gnevyshev_1977, Schove_1979}. \cite{Gnevyshev_1967} states that each 11-year cycle of solar activity consists of two processes with different physical properties. The variety of shapes of the 11-year curves depends on the way these processes overlap. All events in the photosphere, chromosphere and corona, and all kinds of emissions like the radio- and corpuscular emissions take part in these two processes.

It is, nowadays, clear that solar cycle has three phases, an ascending phase, a descending phase and between them Gnevyshev gap (GG) \citep{Storini_2003, Ahluwalia_2004, Bazilevskaya_2006, Norton_2010, Du_2015}, which is kind of a separatrix between the first two (main) phases.  The time of the the Gnevyshev gap is 45-55 months after the start of the nominal cycle length, that is, approximately 33-42\% into the cycle after its start \citep{Takalo_2018, Takalo_2020_2}.

\cite{Storini_2003} gave a review of the earlier studies of the GG effects in different space-weather parameters.
They state that the GG provides a significant time interval in which a relative quietness in the solar terrestrial system ensures no dangerous phenomena in the Earth’s environment. Hence, GG is relevant for the space-Weather and signatures of the GG are present in most of the terrestrial physical parameters.

We study here the SFI using principal component analysis, and show that SFI has a clear Gnevyshev gap (GG) in its first principal component. This gap is deeper than in other solar indices, and is especially clear in the even cycles. We study the mutual strength of the SFI cycles, and the cross-correlation of the geomagnetic disturbance index Ap with SFI. This article is organized as follows: the Section 2 presents the data and methods used in this study. In Section 3 we compare first principal components of selected solar atmospheric indices, and present histograms of different categories of SFI days in Section 4. Section 5 deals with the correlation of SFI and geomagnetic disturbances as measured with Ap-index. In Section 6 we compare the mutual strengths of SFI cycles, and give our conclusions in Section 7.

\section{Data and Methods}

\subsection{Solar Flare Index (SFI)}

The solar flare index used in this study was recently published by \cite{VelascoHerrera_2022}. This database is a composite record of the Astronomical Institute Ond{\v{r}}ejov Observatory of the Czech Academy of Sciences from 1937\,--\,1976 and the records of the Kandilli Observatory of Istanbul, Turkey from 1977\,--\,2020. Although exact total energy of the solar flare event is impossible to determine, the daily and monthly databases are calculated using H$\alpha$ related white-light flares through the formula \citep{Kleczek_1952, Knoska_1984, Ozguc_1989}

\begin{equation}
	Q\,=\,I\times t\:\;     ,
\end{equation}
where {I} is the importance of the flare and t is the duration of the flare in minutes. The importance consists of two factors, the area of the flare (S, 1, 2, 3, 4) and the brilliance of the flare: F(aint), N(ormal) anf B(right) \citep{Ozguc_2002}. The lengths of the solar cycles used in this study are shown in Table 1.

\begin{table}
%\begin{center}
\small
\caption{Sunspot-Cycle lengths and dates [fractional years, and year and month] of (starting) sunspot minima for Solar Cycles 12\,--\,24.} %\citep{NGDC_2013}.}
\begin{tabular}{ c  c  l  c }
  Sunspot cycle    &Fractional    &Year and month     &Cycle length  \\
      number    &year of minimum   & of minimum     &    [years] \\
        \hline   
18    & 1944.1  &1944 February  & 10.2  \\
19    & 1954.3  &1954 April  & 10.5  \\
20    & 1964.8  &1964 October  & 11.7  \\
21    & 1976.5  &1976 June & 10.2  \\
22    & 1986.7  &1986 September  & 10.1  \\
23    & 1996.8  &1996 October  & 12.2  \\
24    & 2009.0  &2008 December  & 11.0  \\ 
25    & 2020    &2019 December 
\end{tabular}
%\end{center}
\end{table}

\subsection{Principal component analysis method}

Principal component analysis is a useful tool in many fields of science including chemometrics \citep{Bro_2014}, data compression \citep{Kumar_2008} and information extraction \citep{Hannachi_2007}. PCA finds combinations of variables, that describe major trends in the data. PCA has earlier been applied, e.g., to studies of the geomagnetic field \citep{Bhattacharyya_2015}, geomagnetic activity \citep{Holappa_2014_2, Takalo_2021_2}, ionosphere \citep{Lin_2012}, the solar background magnetic field \citep{Zharkova_2015}, variability of the daily cosmic-ray count rates \citep{Okpala_2014}, solar corona \citep{Takalo_2022_2} and for separation of the cosmic-ray to solar and Hale cycle related components \citep{Takalo_2022_1}.

In this article we compare, using PCA, solar flare index (SFI) to sunspot numbers (SSN2), solar plage areas (PA), solar 10.7 cm radio flux (RF) and coronal index of solar activity (CI) for Solar Cycles 18\,--\,24. (We omit Cycle 17, because its SFI data is in complete.) To this end, we estimate that the average length of the cycle is 130 months, and use it as a representative Solar Cycle. We first resample the monthly data such that all cycles have the same length of 130 time steps (months), i.e about the average length of the Solar Cycles 18\,--\,24 \citep{Takalo_2018, Takalo_2021_2}. This effectively elongates or abridges the cycles to the same length. Before applying the PCA method to the resampled cycles we standardize each individual cycle to have zero mean and unit standard deviation. This guarantees that all cycles will have the same weight in the study of their common shape. Standardized data are then collected into the columns of the matrix $X$, which can be decomposed as \citep{Hannachi_2007, Holappa_2014_1, Takalo_2018}

\begin{equation}
	X = U\:D\;V^{T}  \     ,
\label{eqn:UDV}
\end{equation}

where $U$ and $V$ are orthogonal matrices, $V^{T}$ a transpose of matrix $V$, and $D$ a diagonal matrix 
	$D= diag\left(\lambda_{1},\lambda_{2},...,\lambda_{n}\right)$
with $\lambda_{i}$ the $i^{th}$ singular value of matrix $X$. The principal component are obtained as the the column vectors of

\begin{equation}
P  = U\!D.
\label{eqn:UD}
\end{equation}
	
The column vectors of the matrix $V$ are called empirical orthogonal functions (EOF) and they represent the weights of each principal component in the decomposition of the original normalized data of each cycle $X_{i}$, which can be approximated as

\begin{equation}
	X_{i} = \sum^{N}_{j=1} \:P_{ij}\:V_{ij} \   ,
	\label{eqn:PV}
\end{equation}

where j denotes the $j^{th}$ principal component (PC). The explained variance of each PC is proportional to square of the corresponding singular value
$\lambda_{i}$. Hence the $i^{th}$
PC explains a percentage
\begin{equation}
\frac{\lambda^{2}_{i}}{\sum^{n}_{k=1}\!\lambda^{2}_{k}} \cdot\:100\%
\end{equation}
of the variance in the data. In this study we use only the first principal component (PC1), which tells the main features of the data and is practically the average of the original data set.

\subsection{Two-Sample T-Test}

The two-sample T-test for equal mean values is defined as follows. The null hypothesis assumes that the means of the samples are equal, i.e. $\mu_{1}=\mu_{2}$. Alternative hypothesis is that $\mu_{1}\neq\mu_{2}$. The test statistic is calculated as
\begin{equation}
T = \frac{\mu_{1} - \mu_{2}}{\sqrt{{s^{2}_{1}}/N_{1} + {s^{2}_{2}}/N_{2}}} ,
\end{equation}
where $N_{1}$ and $N_{2}$ are the sample sizes, $\mu_{1}$ and $\mu_{1}$ are the sample means, and $s^{2}_{1}$ and $s^{2}_{2}$ are the sample variances. If the sample variances are assumed equal, the formula reduces to
\begin{equation}
T = \frac{\mu_{1} - \mu_{2}} {s_{p}\sqrt{1/N_{1} + 1/N_{2}}} ,
\end{equation}
where
\begin{equation}
s_{p}^{2} = \frac{(N_{1}-1){s^{2}_{1}} + (N_{2}-1){s^{2}_{2}}} {N_{1} + N_{2} - 2} .
\end{equation}
The rejection limit for two-sided T-test is $\left|T\right| > t_{1-\alpha/2,\nu}$, where $\alpha$ denotes significance level and $\nu$ degrees of freedom. The values of $t_{1-\alpha/2,\nu}$ are published in T-distribution tables \citep{Snedecor_1989, Krishnamoorthy_2006, Derrick_2016}.  Now, if the value of p$<\alpha=0.05$, the significance is at least 95\%, and if p$<\alpha=0.01$, the significance is at least 99\%.

\subsection{Cross-correlation}

The cross-correlation function measures similarity between a time series and lagged versions of another time series as a function of the lag. Let us consider two time series (vectors) $x_{t}$ and $y_{t}$,. We define the cross-covariance as \citep{Box_2016}

\begin{equation}
 c_{xy}(k) = \frac{1}{T}  \displaystyle\sum^{T-k}_{t=1}\left(x_{t}-\overline{x}\right)\!\left(y_{t+k}-\overline{y}\right), k=0,1,2,...
\end{equation}
and
\begin{equation}                          
 c_{xy}(k) = \frac{1}{T}  \displaystyle\sum^{T+k}_{t=1}\left(y_{t}-\overline{y}\right)\!\left(y_{t-k}-\overline{y}\right          ), k=0,-1,-2,...
\end{equation}
where $\overline{x}$ and $\overline{y}$ are sample means of the time series. If we use only part of the series in calculation, we call the procedure sample cross-covariance. The sample standard deviations are $s_{x}=\sqrt{c_{xx}(0)}$
and  $s_{x}=\sqrt{c_{yy}(0)}$. As an estimate of the sample cross-correlation we have
\begin{equation}
r_{xy}(k) = \frac{c_{xy}(k)}{s_{x}s_{y}} ,  k=0, \pm1, \pm2,...
\end{equation}

We calculate here the cross-correlation of two time series in two ways, i.e. from a limited sample of the time series and using circular cross-correlation of the time series through the whole solar cycle. The fastest way to do the process is by using Fourier-transforms of the vectors \textbf{x}, \textbf{y}, multiply them (no conjugate is needed here, because the vectors are real valued) and take an inverse Fourier transform of the product. The result can be normalized by dividing with the norms of the vectors.

\section{PC Analyses of Selected Solar Atmospheric Indices}
	\label{PCA}

Principal component analysis (PCA) is a good method to distinguish essential features in the time series. This method is especially good for analysis of time series, which depend on a single strong period, e.g. solar cycle. This is because, it is probable that this period forms the base of the first principal component (PC1). Because PCs are orthogonal, the second component is most likely related to the period, which is half of the solar period. Now if the PC2s of the successive cycles are in opposite phase we get twice the original period, i.e Hale period, at least in the case the time series in question is also depending on the Hale cycle. (Notice that normalized solar cycle is actually one period of a sine wave, and we know that $sin(k \alpha)$ and $sin(n \alpha)$ are orthogonal, when $k\neq n$.) In this way PCA is better here than e.g. Independent Component Analysis (ICA), which finds components that are maximally independent from each other, e.g. noise from the signal (ICA components are not necessarily orthogonal). \cite{Takalo_2021_2, Takalo_2022_1} has actually shown, using PCA, that the aa-index and cosmic-ray indices can be separated to solar cycle and Hale cycle related components.
 
\begin{figure} 
 \centerline{\includegraphics[width=1.0\textwidth,clip=]{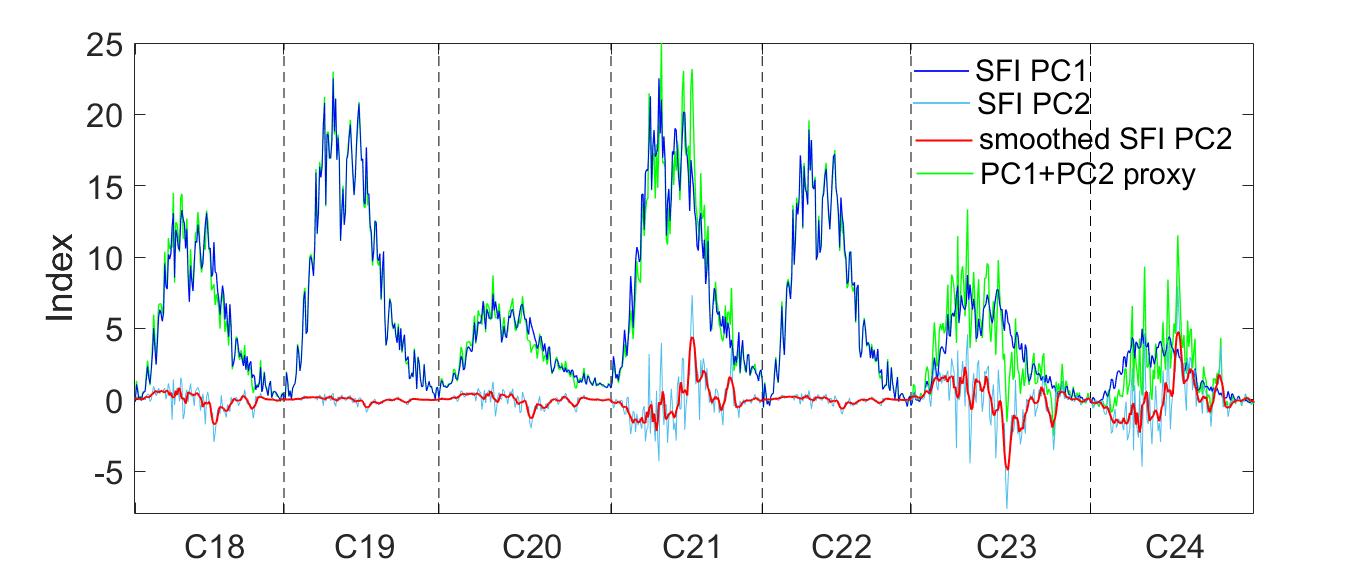}}
 \caption{a) The PC1, PC2, smoothed PC2 and the PC1+PC2 time series of the SFI for the Solar Cycles 18\,--\,24.}
 \label{fig:PC_proxies}
 \end{figure}

Here we are interested mostly the PC1 of SFI and compare it to other PC1s of solar cycle related indices. Because the GG is known as an essential feature of the solar cycle, it is also usually present in the PC1, which is the component accounting for the most variance of the data. That is why we conducted PCA for several solar indices. Because PCA is a matrix-based method, we use the procedure described earlier to have all Cycles 18\,--\,24 the same length 130 time steps (months), which is about the average length of those cycles. The idea in this method is that we suppose all cycles to have the same dynamical phases, except that cycles differ in the duration of the phases. The separate Cycles 18\,--\,24 form the columns of the matrix $X$ in the Eqs. \ref{eqn:UDV}--\ref{eqn:PV}. The first and second principal components of each cycle i are $X_{i}$s with j=1 and j=2, respectively in the Eq. \ref{eqn:PV}. We then return each cycle back to its original length, and then back to its original amplitude by multiplying PCs of each cycle with the standard deviation of the original cycle and and adding the mean value of the original cycle to PC1. Now we can construct full PC1 and PC2 time series by concatenating the cycles to their original order. Figure \ref{fig:PC_proxies} shows time series the PC1 and PC2 series and their sum time series PC1+ PC2 of the SFI for Solar Cycles 18\,--\,24. Note how every cycle has evidently the same gap in their same PC1. Some cycles have, however, very large fluctuation in its PC2, which explains 12.4\,\% of the variance in the solar cycle time series. This is the case especially for Cycle 21, 23 and 24. For Cycle 23 the PC2 lowers the GG, but makes it still deeper for Cycle 21 and 24. It should be noted that there exists PC3, which accounts for 8.2\.\% of the variance (not shown here), and is especially large for Cycles 20 and 23. Other PCs are quite negligible.

\begin{figure} 
 \centerline{\includegraphics[width=1.05\textwidth,clip=]{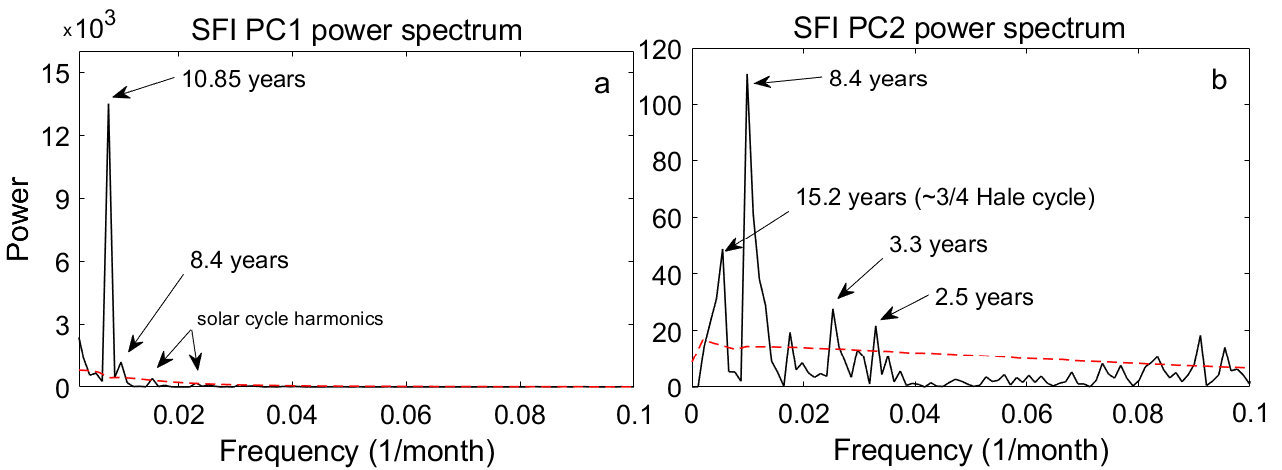}}
 \caption{a) The power spectrum of SFI PC1 time series for SC18\,--\,SC24. b) a) The power spectrum of SFI PC2 time series for SC18\,--\,SC24.}
 \label{fig:PC_power_spectra}
 \end{figure}

Figure \ref{fig:PC_power_spectra}a, b and show the power spectra of the PC1 and PC2 time series for the Solar Cycles 18\,--\,24, respectively. We have low-pass filtered the time series, because we are interested here only multiple year periods in the data. The dashed red lines are 99\,\% significance levels of the red-noise data. Red-noise is calculated as a power spectrum from the time series
\begin{equation}
	x_{n} = \alpha\:x_{n-1}+z_{n}  ,
\end{equation} 
where $\alpha$ is lag-1 autocorrelation and $z_{n}$ is Gaussian white noise. Furthermore, we adjust the length and variance of the red-noise time series to those of the original PC1 or PC2 time series \citep{Torrence_1998, Mendoza_2011, Oloketuyi_2019}. There are other ways to approximate the significance level, e.g. so-called false alarm probability (FAP) method \citep{Ozguc_2003}. We, however, think that it is not the best way in our case to find the confidence limit. This is because we have separated different periods into several different orthogonal time series. 
Note that most of the earlier studies have concentrated on the short- and mid-term periods in the solar flares as discussed in the Introduction. It is understandable that by far the strongest power in PC1 is at 10.85 years (130 months), which is the average length of the Solar Cycles 18\,--\,24. Another peak above the red-noise level in the PC1 time series is the 8.4-year period. The rest peaks are negligible and the two peaks marked in the Figure \ref{fig:PC_power_spectra}a are periodic peaks of the solar cycle. It is interesting that the 8.4-year peak is also present and actually the most powerful peak in the power spectrum of the PC2 time series (Figure \ref{fig:PC_power_spectra}b). The other peaks above the 99\,\% confidence level are about 15.2-, 3.3- and 2.5-year periods. Note, that the PC2 power spectrum does not show the solar cycle period. Because the resolution for the power spectrum of quite short time series (monthly data) is quite poor, we suppose that the periods 3.3 and 8.4 add together the solar cycle (note that these peaks are skewed to the right, which means that the exact period is somewhat less than the aforementioned values). On the contrary the 15.2-year peak is skewed to the left, and should probably be slightly longer period. We believe that this peak is 3/4 of the magnetic 22-year Hale cycle.
\cite{VelascoHerrera_2022} and \cite{Ozguc_2021} reported in their research about 3.5-year, 3.2-year, respectively. These are quite near to our 3.3-year period in PC2 power spectrum. It should be noted that we get 3.16- and 1.46-year peaks for the power spectrum of the PC3 time series (not shown here). These peaks may be related to the 1153-day peak in total solar surface and about 540-day peak in the northern and southern hemispheres
of the Sun reported by \cite{Ozguc_2003}.

\begin{figure} 
 \centerline{\includegraphics[width=1.0\textwidth,clip=]{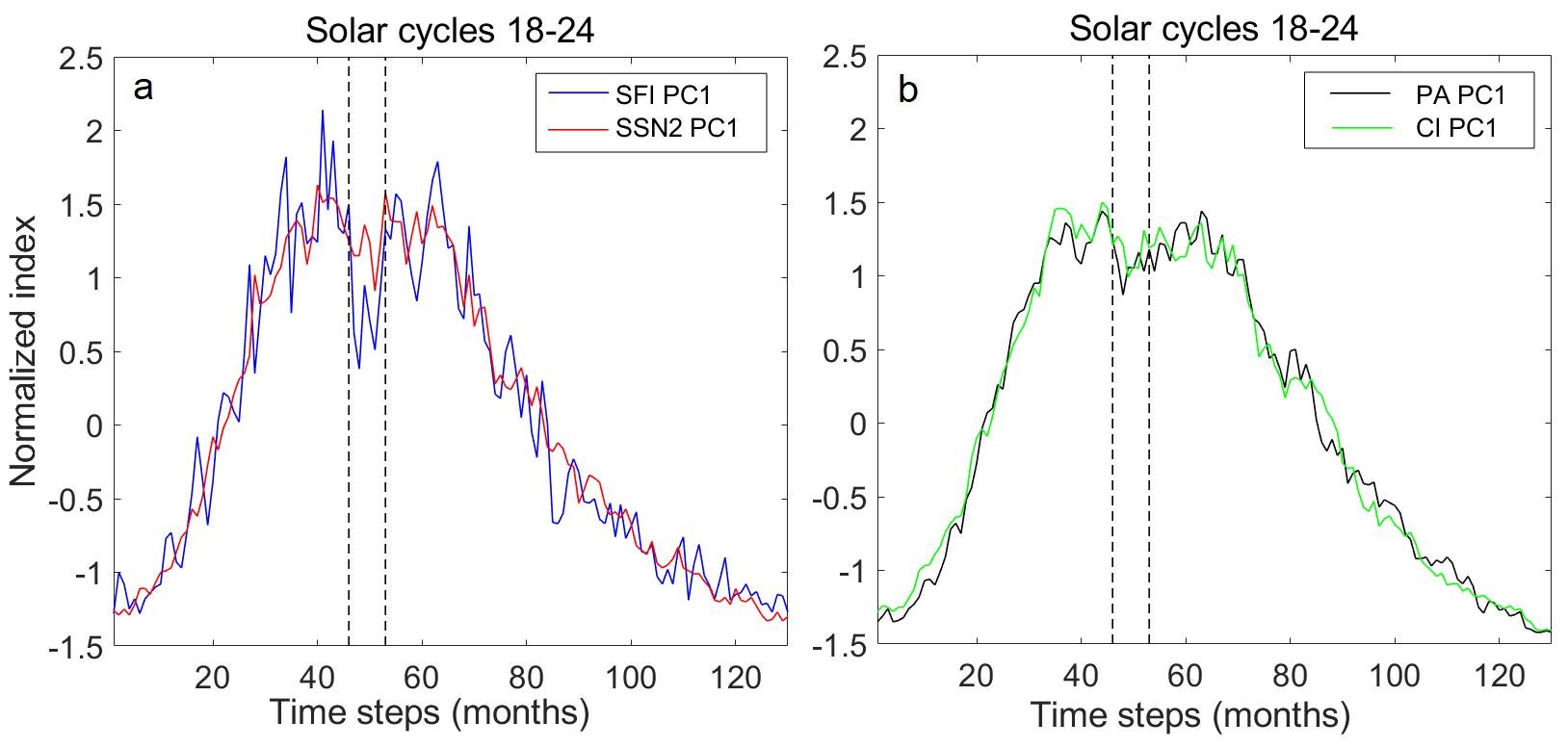}}
 \caption{a) The PC1s of solar flare index (SFI) and sunspot numbers (SSN2) b) The PC1s of plage area (PA) index and solar corona index (CI).}
 \label{fig:PC1s}
 \end{figure}

\begin{figure} 
 \centerline{\includegraphics[width=1.0\textwidth,clip=]{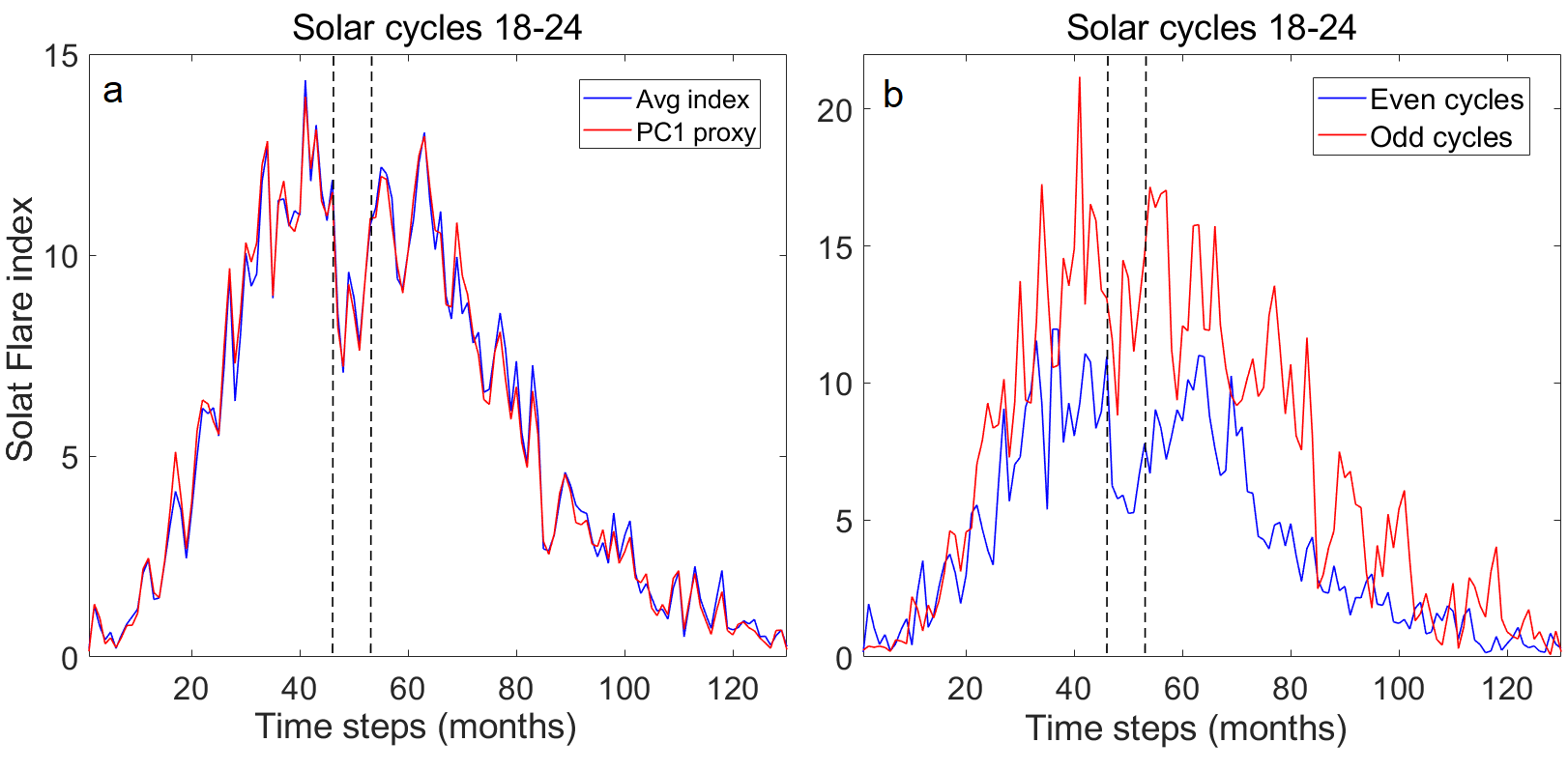}}
 \caption{a) The PC1 and average cycle of SFI for the Cycles 18\,--\,24. b) The average cycle for the even and odd Cycles 18\,--\,24.}
 \label{fig:PC1_and_Avg}
 \end{figure}

Figure \ref{fig:PC1s}a and b show the first principal components (PC1) for four solar atmospheric indices for Cycles 18\,--\,24, i.e. for sunspot number SSN2 (photosphere), solar flare index (SFI) (chromosphere), plage area (PA) (chromosphere), 10.7 cm radio flux (RF) (high chromosphere, low corona),  and solar green line corona index (CI), which is actually measured at low altitude (about (60 arcsecs) above the surface of the Sun. It is surprising, that SFI has most conspicuous gap between 46 to 53 (shown as vertical dashed lines) months in its PC1. 
Although PC1 for SFI explains just 62.7\,\% of the total variance of the data, it is, as said earlier, practically the average of the Cycles 18\,--\,24. This is seen in Fig. \ref{fig:PC1_and_Avg}a, which shows the PC1 reverted back to original amplitude and the average of the Cycles 18\,--\,24. Note that the gap, which we believe is the Gnevyshev gap, has two-fold minimum with a peak between. The bottom of the first minimum at 48 months is 40\,\% smaller than the peak two months earlier. It turns out that the first minimum is due to both even and odd cycles and the second minimum mainly to even cycles (see Fig. \ref{fig:PC1_and_Avg}b). Note also that the drop for the even cycles is about 50\,\%.

\begin{figure} 
 \centerline{\includegraphics[width=1.0\textwidth,clip=]{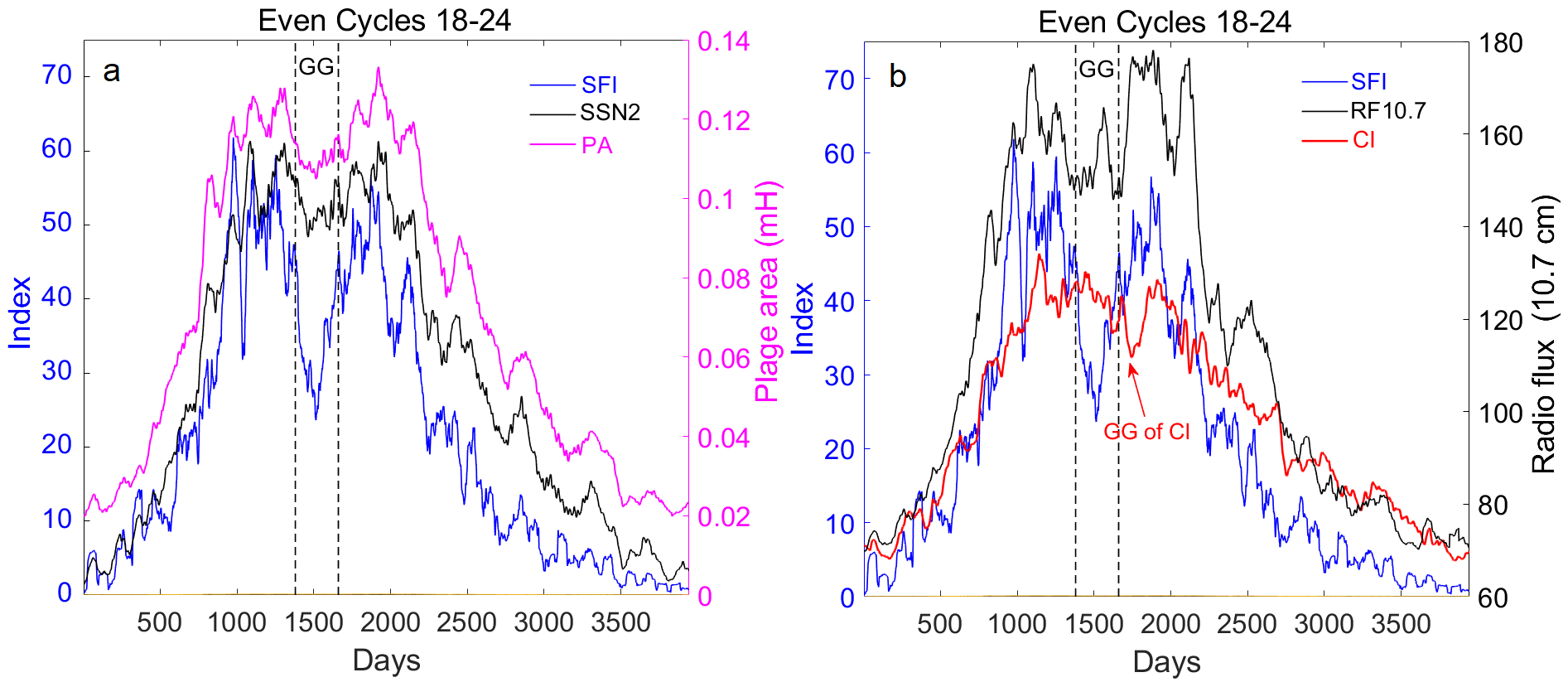}}
 \caption{a) The total daily index of SFI, SSN2 and PA for the even Cycles 18\,--\,24. b) The total daily index of SFI, CI and RF for the even Cycles 18\,--\,24. (The indices are smoothed over 31 days or 61 days and SSN2 is adjusted to enable presenting with the same axis as other indices.)}
 \label{fig:Indices_even_3945}
 \end{figure}

\begin{figure} 
 \centerline{\includegraphics[width=1.0\textwidth,clip=]{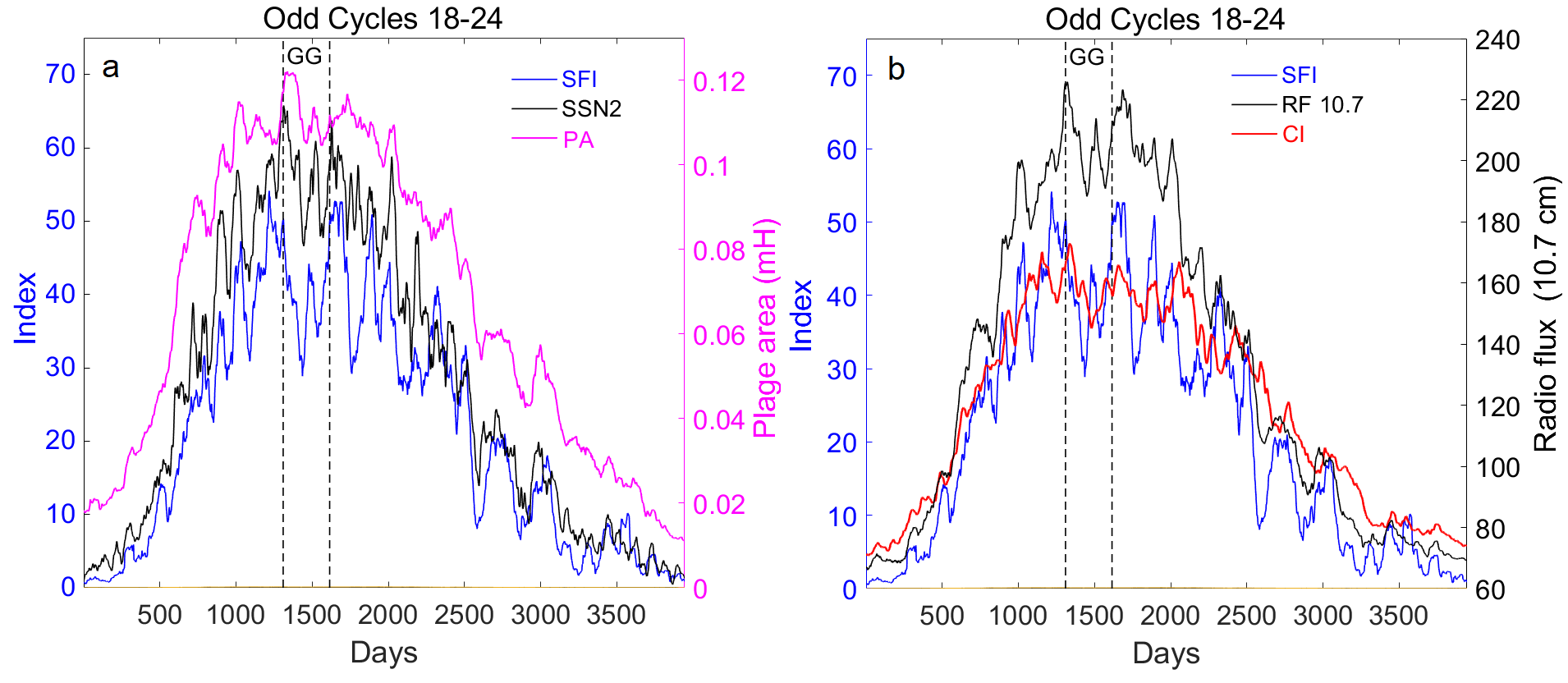}}
 \caption{a) The total daily index of SFI, SSN2 and PA for the odd Cycles 18\,--\,24. b) The total daily index of SFI, CI and RF for the odd Cycles 18\,--\,24. (The indices are smoothed over 31 days or 61 days and SSN2 is adjusted to enable presenting with the same axis as other indices.)}
 \label{fig:Indices_odd_3945}
 \end{figure}

In order to study the GG in more detail, we use daily data of the aforementioned indices. We have interpolated all cycles to have 3945 days, which is about the average number of days in the Cycles 18\,--\,24. Figure \ref{fig:Indices_even_3945}a and b show the total amount of SFI, SSN2, PA and SFI, CI, RF for the even Cycles 18\,--\,24, respectively. Note that SSN2 has been adjusted to enable the presentation of the indices with the same axis. Furthermore, all indices are smoothed over 31 or 61 days (depending how spiky they are in order to make the figure more readable). Note that the GG locates similarly for SFI, SSN2 and PA in Fig. \ref{fig:Indices_even_3945}a, but SFI has by far deepest decline between the dashed vertical lines, which are 1380 and 1660 days after the minimum of the cycles, i.e. show about a nine month window. The SFI and RF are also simultaneous in Fig. \ref{fig:Indices_even_3945}b, except that RF is twofold with a peak in between the GG region. The GG in the CI locates, however, later but is twofold like the GG of the RF index. Figure \ref{fig:Indices_odd_3945}a and b show the total amount of SFI, SSN2, PA and SFI, CI, RF for the daily odd Cycles 18\,--\,24. Note again that the decline of SFI for the odd cycles is not as deep as for the even cycles. The clearest difference here for the even cycles in Fig. \ref{fig:Indices_even_3945} is that all GGs have two minima with a peak in between. The GG window is also somewhat earlier for the odd cycles compared to the even cycles, i.e. between 1310 and and 1615 days after the minimum. It also seems that the GG starts somewhat later for all other indices than the SFI for the odd cycles. Interestingly as seen in Fig. \ref{fig:Indices_odd_3945}a the GG for SSN2 starts about one solar rotation later than the decline for the SFI. Note also from Fig. \ref{fig:Indices_odd_3945}b that the start of the decline of the RF and CI is simultaneous for the odd cycles and also about one solar rotation later than the decline for the SFI. These lags may be due to long-living recurrent sunspot groups \citep{Nagovitsyn_2019}. On the other hand, the GG of the PA starts much later, and actually at about the time, 1380 days, marked in the Fig. \ref{fig:Indices_even_3945} as a starting point of the GG window.

\section{Histograms of the daily SFIs}
		\label{Hist} 

\begin{figure} 
 \centerline{\includegraphics[width=1.0\textwidth,clip=]{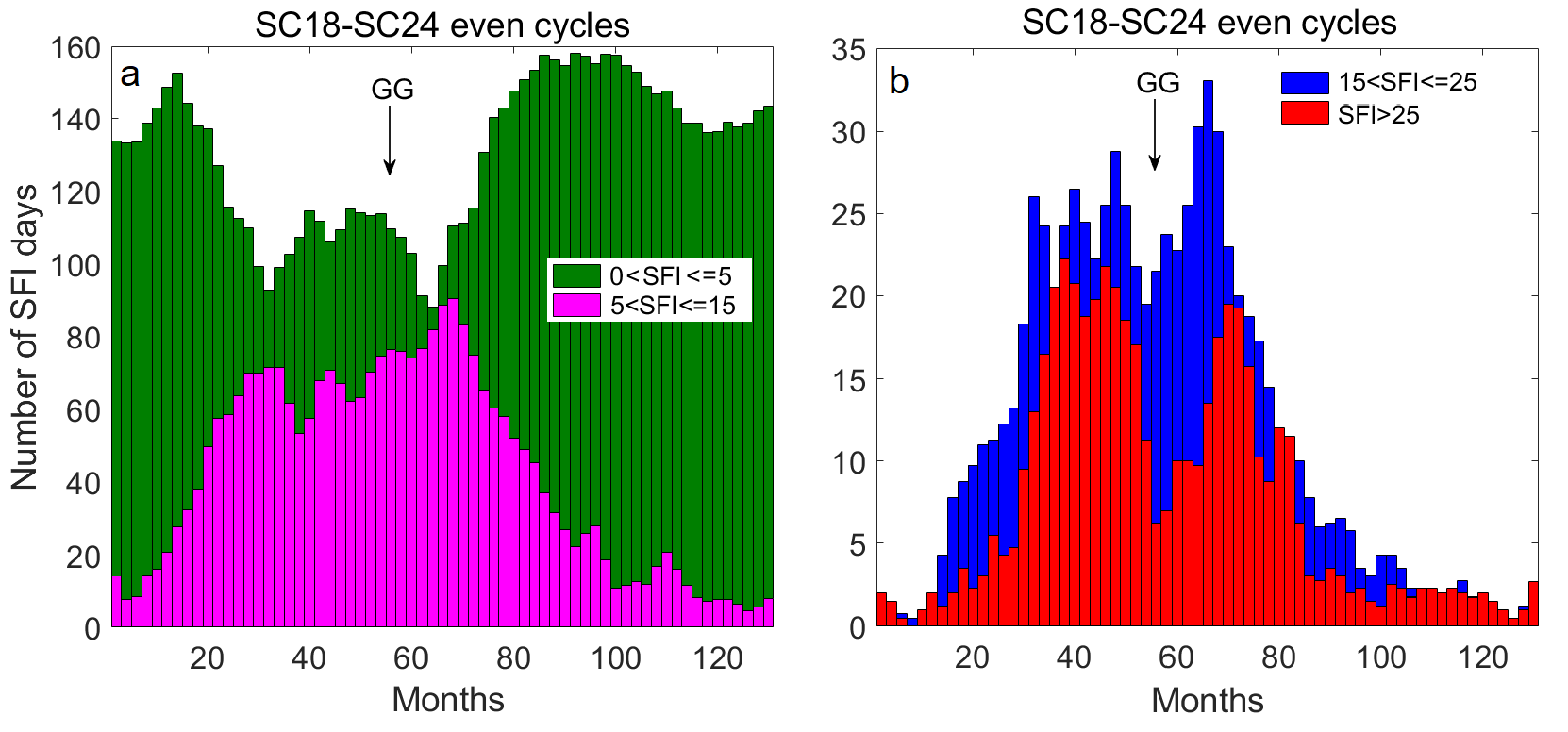}}
 \caption{The histograms of amounts for SFI days a) in categories weak (0$<$SFI$\leq$5) and moderate (5$<$SFI$\leq$15), and b) in categories strong (15$<$SFI$\leq$25) and very strong (SFI$>$25) for the even Cycles 18\,--\,24.}
 \label{fig:categories_even}
 \end{figure}

\begin{figure} 
 \centerline{\includegraphics[width=1.0\textwidth,clip=]{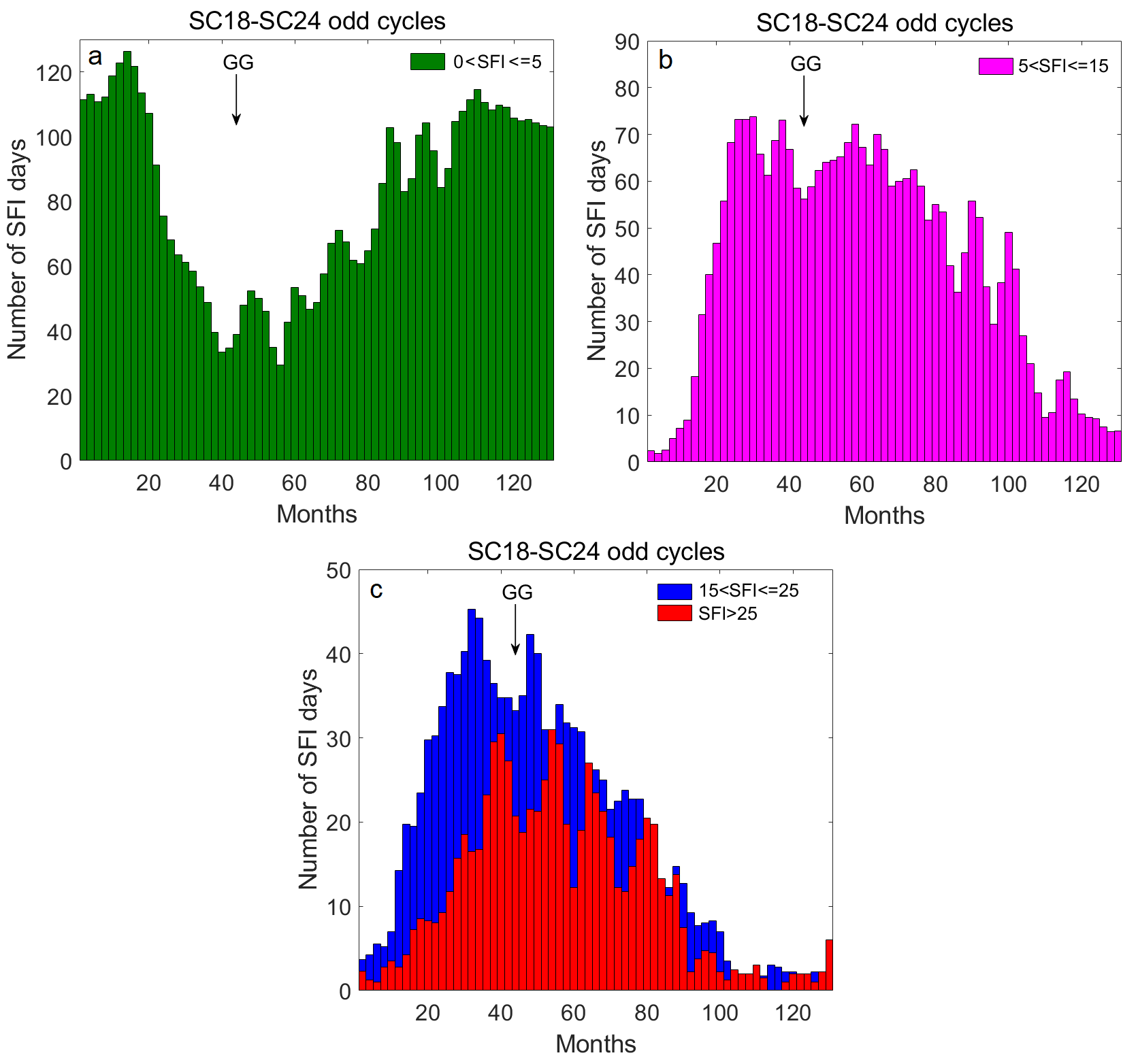}}
 \caption{{The histograms of amounts for SFI days a) in category weak (0$<$SFI$\leq$5) b) in category moderate (5$<$SFI$\leq$15), and c) in categories strong (15$<$SFI$\leq$25) and very strong (SFI$>$25) for the odd Cycles 18\,--\,24.}}
 \label{fig:categories_odd}
 \end{figure}

The huge GG in the average cycle of SFI waked our interest in more detailed analysis of the gap. We divide the daily values of SFI to four categories: weak (0 $<$SFI$\leq$ 5), moderate (5 $<$SFI$\leq$ 15), strong (15 $<$SFI$\leq$ 25) and very strong (SFI $>$ 25). Figures \ref{fig:categories_even} and \ref{fig:categories_odd} show the histograms of the categories for the even and odd cycles 18\,--\,24, respectively. Each bar represents daily values of two months. Because the SFI data is quite spiky, we smoothed the data over six months. Looking at the figures, it is evident that odd cycles have more strong and very strong category SFI days at the maximum of the cycles, and consequently deeper valley in the histogram of the weak category days. Note that there is a hump at the GG site in the histogram of the weak category. Both even and odd cycles have about the same amount moderate category SFI days. The even cycles seem to have deeper GG (shown with an arrow) in strong and especially very strong category days. Note that, consequently, there are more moderate and weak category days during the gap in the even cycles. On the contrary, the GG in the odd  cycles exists already in the moderate category histogram, and also in the strong and very strong category histograms. The difference of the odd cycles compared to even cycles is that the gap is not very deep in any of these categories. The GG is also somewhat earlier for the odd cycles than the even cycles. These results are consistent with the earlier studies by \cite{Takalo_2020_1, Takalo_2020_2} concerning the GG for the sunspot numbers and sunspot groups.

\section{The response of IMF Bv2 and Ap-index to SFI}
	\label{Resp}

\begin{figure} 
 \centerline{\includegraphics[width=0.5\textwidth,clip=]{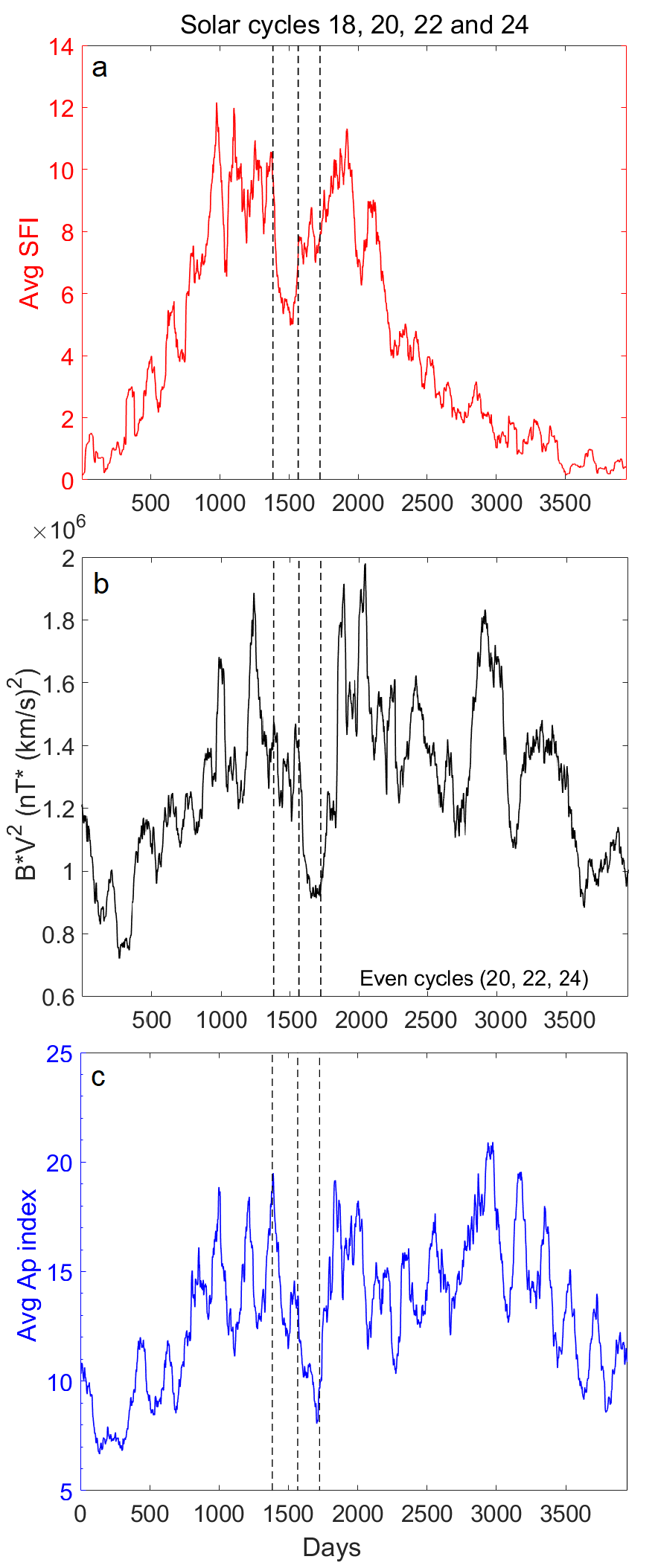}}
 \caption{Top panel: The average solar flare index for the even Solar Cycles 18\,--\,24. Middle panel: The average IMF$Bv^{2}$-component for the even Cycle 20\,--\,24. Bottom panel: The average Ap-index for the even Solar Cycles 18\,--\,24.}%
 \label{fig:SFI_Bv2_Ap_even}
 \end{figure}

To show the response of IMF and geomagnetic field to SFI, we reconstruct 3945 day long data of SFI for the even Cycles 18\,--\,24, daily Ap-data for even Cycles 18\,--\,24 and daily interplanetary magnetic field/solar wind function $Bv^{2}$ \citep{Ahluwalia_2000} for even Cycles 20, 22 and 24. (Note that we have IMF data only for Cycles 20\,--\,24.) We use 3945 days as an average length of the cycle, because it is about 130 months, which we used earlier in the monthly analyses (see also \citep{Takalo_2021_1}). Figure \ref{fig:SFI_Bv2_Ap_even}a, b and c show the sequence of daily averages of SFI, $Bv^{2}$ and Ap-index for the even Cycles between 18\,--\,24 (except 20\,--\,24 for $Bv^{2}$), respectively. Here the indices are smoothed using trapezoidal smoothing over 61 points (rectangular moving-average smoothing with window end points having only half of the weight of the inner points). The vertical dashed lines are at 1384, 1567 and 1725 days from the start of the Cycle. The decline of the daily SFI values (Fig. \ref{fig:SFI_Bv2_Ap_even}a) between lines 1384\,--\,1725 (mean=6.70) is at least 99\,\% significant in two-sample T-test for unequal mean compared to similar intervals earlier (p$<10^{-11}$, mean=9.66) and after (p$<10^{-10}$, mean=8.96) as calculated from the unsmoothed data. Note, however, that the deepest phase of the decline between 1384 and 1567 days lasts about 6-months, which we suppose to be the main GG phenomenon. Note also a quasi-periodic structure in the average SFI. The mean period of this fluctuation is about 150 days, which has been found to be the period of various activities of the Sun \citep{Rieger_1984, Lou_2000, Richardson_2005, Takalo_2021_1, Li_2021, VelascoHerrera_2022}.

The decline in IMF $Bv^{2}$ near the Earth (Fig. \ref{fig:SFI_Bv2_Ap_even}b) and Ap-index (Fig. \ref{fig:SFI_Bv2_Ap_even}c) is not simultaneous with the drop in the SFI. There is, however, a deep prolonged decline, which lasts, at least until day 1725. Note the similar shape of the IMF $Bv^{2}$ behavior between 1567\,--\,1725 days compared to SFI between 1384\,--\,1567 days. If we assume that this part of the decline (which is also related to other solar events) travels with the speed of solar wind, the lag is quite near to six months, i.e. about same as the lag from the drop in SFI at 1384 to drop in IMF $Bv^{2}$ and Ap at 1567 days. It should be noted, that geomagnetic disturbances (or at this case absence of disturbances) usually lag the solar indices some months. For example, \cite{Takalo_2021_2} has estimated that aa minima for Cycles 10\,--\,24 lag the sunspot number minima from 3 (Cycle 14) to 17 (Cycle 16) months, i.e. 10 months on the average.

\begin{figure} 
 \centerline{\includegraphics[width=1.0\textwidth,clip=]{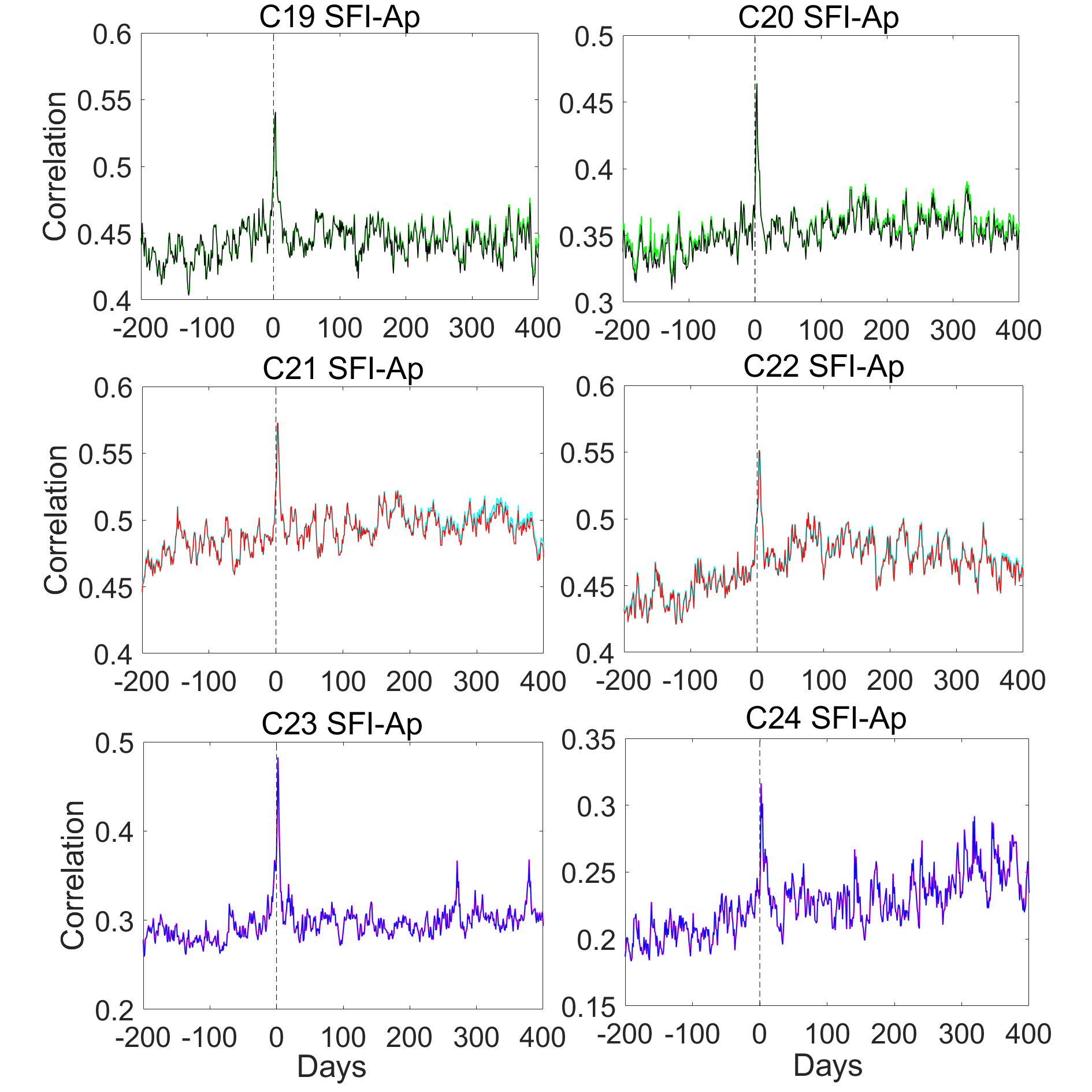}}
 \caption{The cross-correlations between SFI and Ap-index for for the Solar Cycles 18\,--\,24.}
 \label{fig:SFI_Ap_corr}
 \end{figure}

\begin{figure} 
 \centerline{\includegraphics[width=1.0\textwidth,clip=]{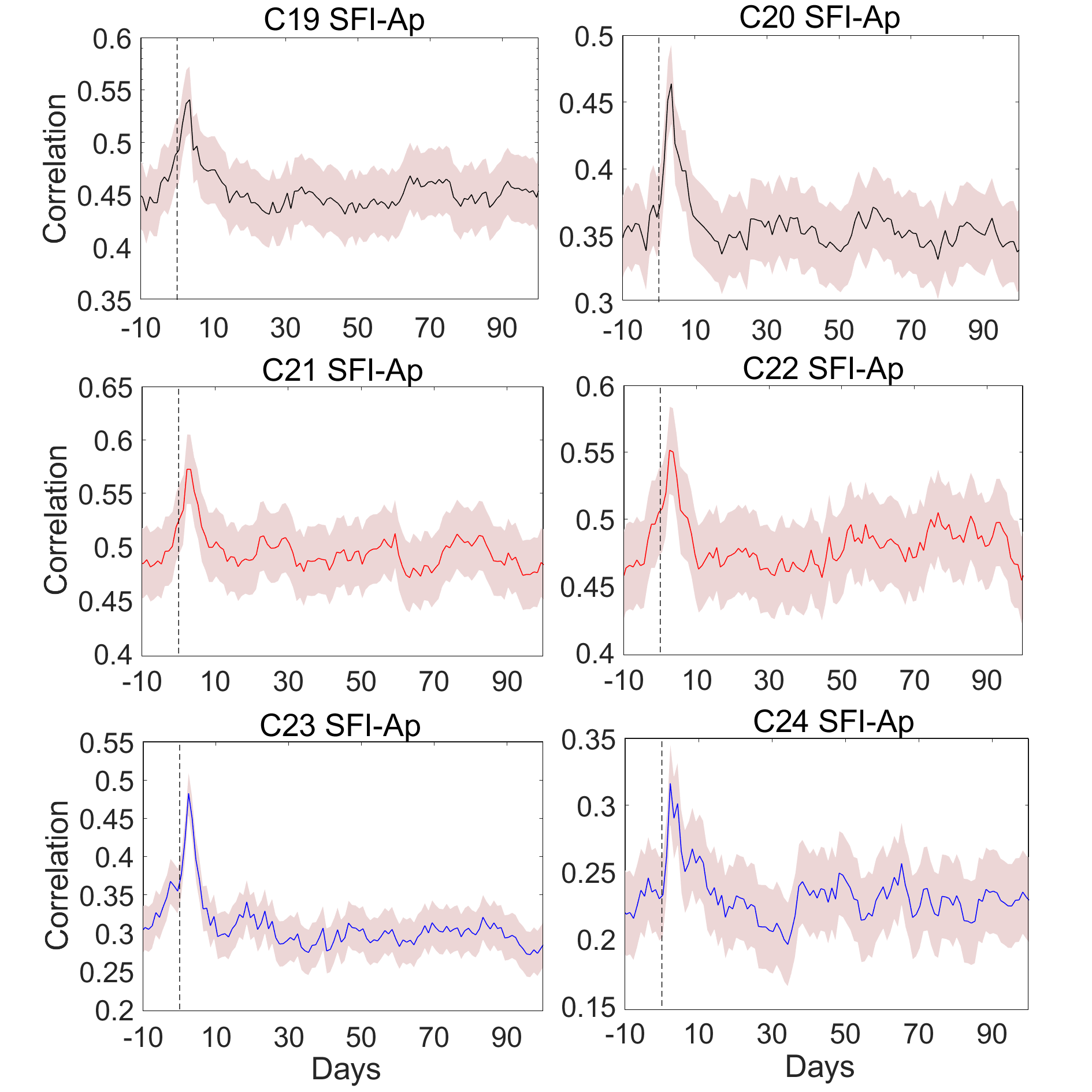}}
 \caption{Magnification of the cross-correlations between SFI and Ap-index for for the Solar Cycles 18\,--\,24. The lilac color in the background shows two standard deviation limits.}
 \label{fig:SFI_Ap_corr_magn}
 \end{figure}

Another case is the instantaneous response of the geomagnetic field to the powerful events like CME/shock waves, which may have transit times measurable from hours to a few days \citep{Suresh_2022, Cliver_2022}. These events are often related to H$\alpha$ flares. Figure \ref{fig:SFI_Ap_corr} shows the cross-correlation of SFI and Ap-index for the Solar Cycles 19\,--\,24. Negative days are SFI Ap preceding the SFI and positive days Ap succeeding the SFI. There is a peak maximizing at minus two to three days in all cross-correlation curves. Note also that cross-correlation stays somewhat higher after the peak. The black (uppermost panels), red (middle panels) and blue (lowest panel) are calculated as a limited interval around the zero point, and The green (uppermost panels), cyan (middle panels) and magenta (lowest panel) are calculate using circular cross-correlation. The overall level of the cross-correlation seems to be proportional to the strength of the corresponding cycle, although not exactly. The highest peak above the overall level is for the Cycle 23, which may due to some strong flares in the beginning of the new millennium. Figure \ref{fig:SFI_Ap_corr_magn} shows the magnifications of the Fig. \ref{fig:SFI_Ap_corr} with two standard deviations as a lilac color. Note that the standard deviation is narrowest for aforementioned Cycle 23.

\begin{figure} 
 \centerline{\includegraphics[width=1.05\textwidth,clip=]{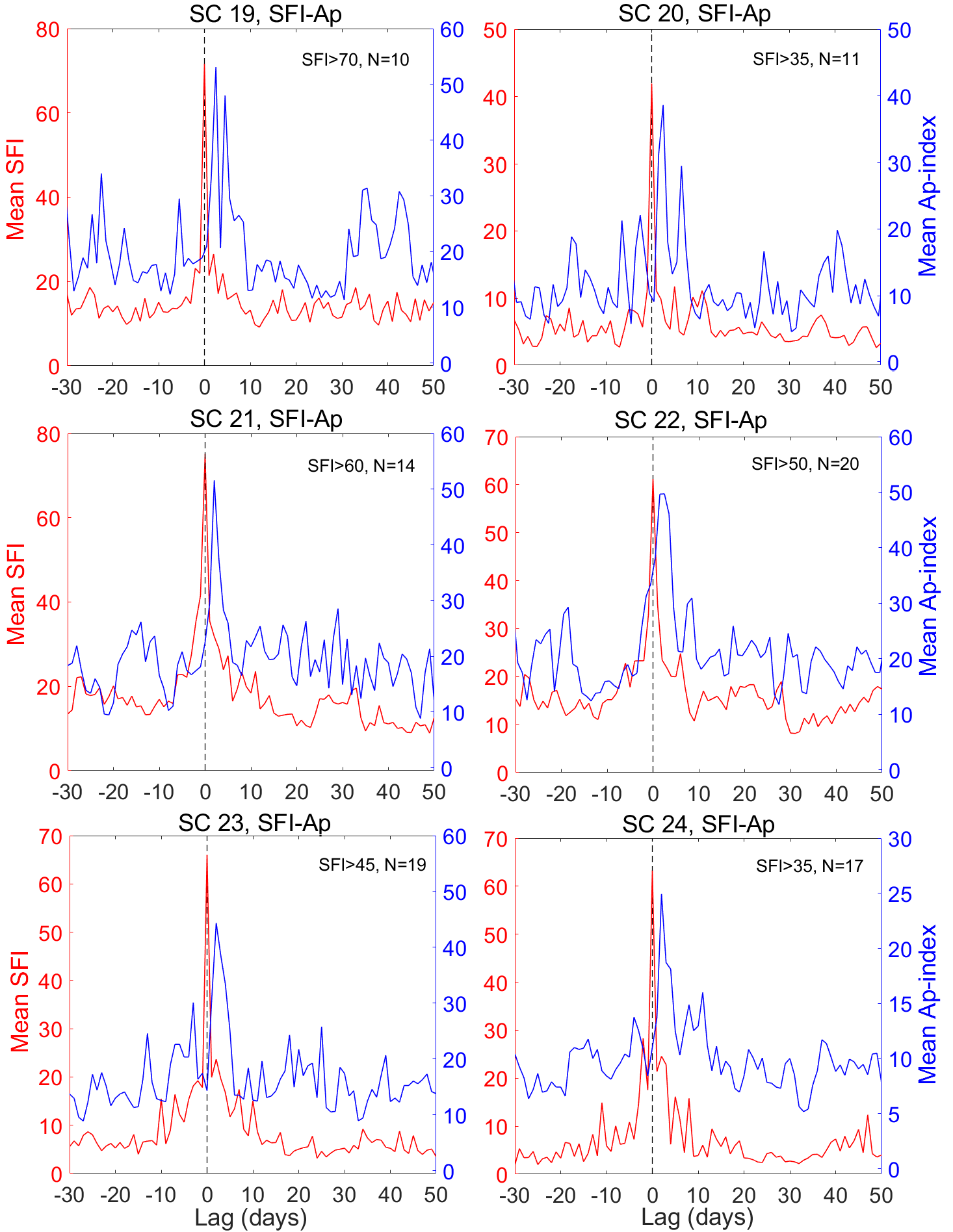}}
 \caption{Superposed-epoch analyses of SFI\,--\,Ap mutual dependence for the Solar Cycles 19\,--\,24.}
 \label{fig:SFI_Ap_epoch_analyses}
 \end{figure}

Another way to study the immense causal response of Ap-index to SFI is to use the so-called superposed-epoch analysis \citep{Kharayat_2016, Pokharia_2018}. This is actually similar to what we have used earlier in our PC analysis in Section \ref{PCA} and in the Figure \ref{fig:SFI_Bv2_Ap_even}, i.e. superposed the cycles \citep{Takalo_2018, Takalo_2021_1, Takalo_2021_2, Takalo_2022_1}. The difference is here that we use another time series (signal) as a trigger in recording the epochs. Here we use as a zero day the timestamp when SFI exceed some fixed level, which depends on the activity of the cycle. We choose the level such that one cycle includes 10-25 epochs. Figure \ref{fig:SFI_Ap_epoch_analyses} shows the superposed epoch analyses for the Solar Cycles 19\,--\,24. The zero day shows the average SFI (red), which has triggered the recording, and blue curve shows the response of the Ap-index to the SFI. Note that Ap maximizes 2-3 days after the triggering peak in SFI. It is evident that the results are similar to the cross-correlation analyses. We show here the time interval -30\,--\,50 days, i.e start recording 30 days before the zero day. The inserted text in the figures shows the SFI triggering level and the number of superposed epochs in each cycle. Note that cycles 19 and 20 differ from others such that there are two peaks after the peak in SFI. This is probably related to the knee in the peaks of the Cycles 19 and 20 in Figure \ref{fig:SFI_Ap_corr_magn}. Furthermore, Cycle 24 has a twofold peak similarly to that in the cross-correlation of Cycle 24 in Figure \ref{fig:SFI_Ap_corr_magn}.

\section{Comparison of the SFI Cycles}

\begin{figure} 
 \centerline{\includegraphics[width=1.05\textwidth,clip=]{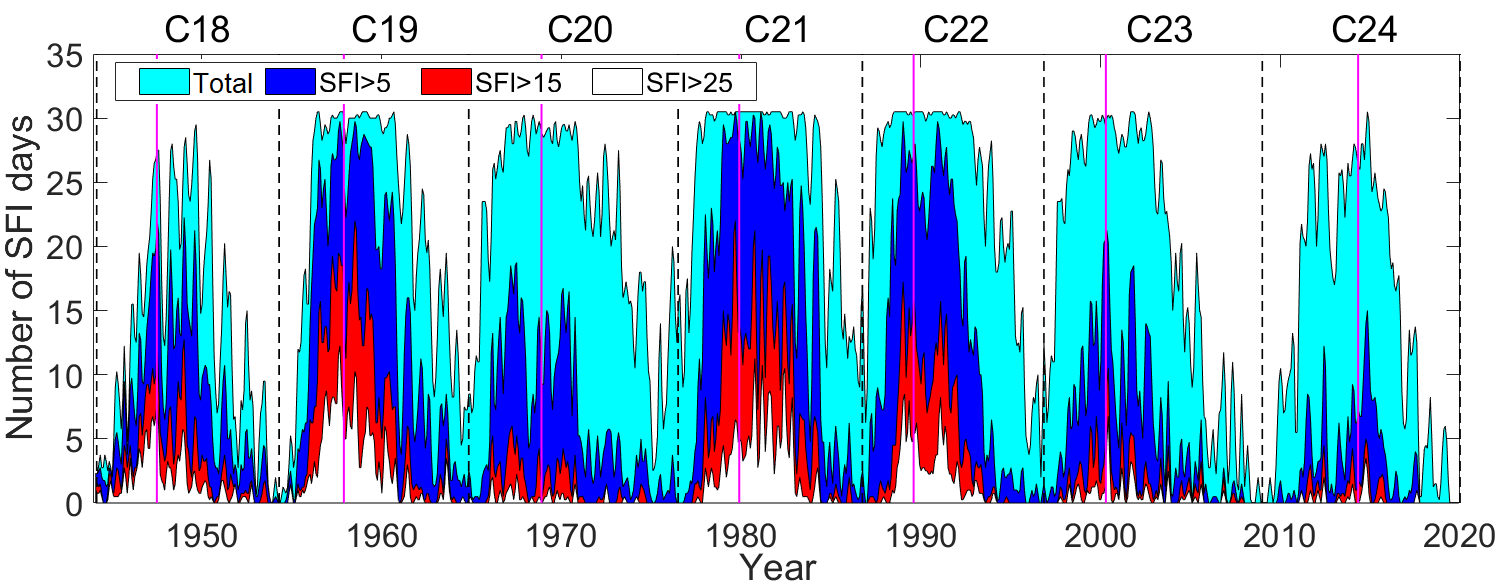}}
 \caption{Number of SFI days as cumulative categories for the Solar Cycles 18\,--\,24.}
 \label{fig:SFI_Cycles}
 \end{figure}

Figure \ref{fig:SFI_Cycles} shows the different categories of the number of SFI days for the Solar Cycles 18\,--\,24. The figure is cumulative such that the cyan color shows all days with SFI $>$0, blue color SFI $>$5, red color SFI $>$15 and white color SFI $>$25. Everything under y-axis value 30 and above the cyan color area means days with no H$\alpha$ solar flare. Note that saturation level is at 30, because each point of the curves represents one month. (We have used three month trapezoidal smoothing to make figure more readable, and the small lumps on top of the saturation level of the cyan curve are due to months which have 31 days.) The minima between the cycles are shown as dashed black vertical lines and the maxima of the cycles with magenta vertical lines.

\begin{figure} 
 \centerline{\includegraphics[width=0.9\textwidth,clip=]{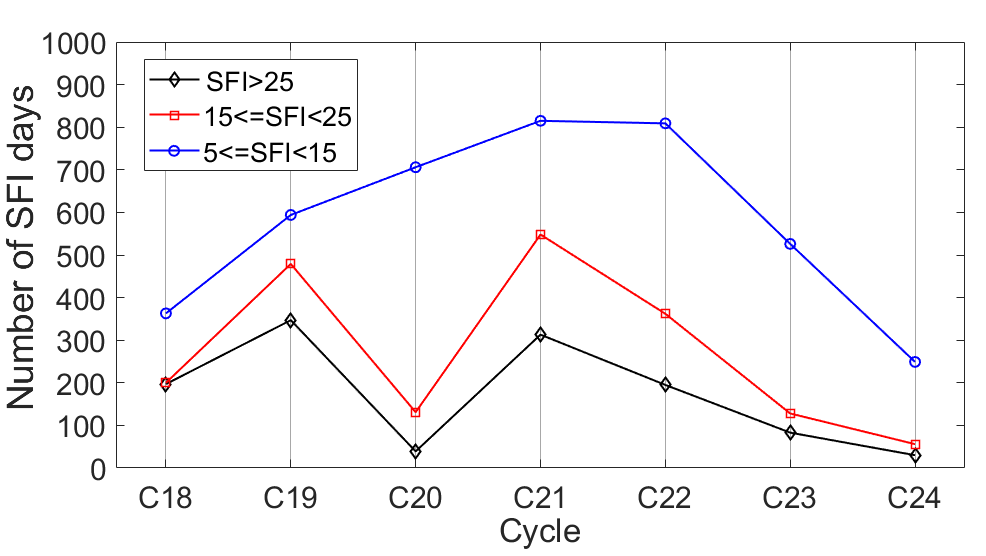}}
 \caption{Number of SFI days for moderate (5$<$SFI$\leq$15), strong (15$<$SFI$\leq$25) and very strong (SFI$>$25) categories of the Cycles 18\,--\,24.}
 \label{fig:SFI_Numbers}
 \end{figure}

Figure \ref{fig:SFI_Numbers} shows the different categories of the solar flare days for the Solar Cycles 18\,--\,24 in a different way, i.e. as total numbers of different category, moderate (5$\leq$SFI$<$15), strong (15$\leq$SFI$<$25) and very strong (SFI$>$25) SFI days. Cycles 19, 21 and 22 seem to have H$\alpha$ flares almost every day, except at the beginning and end of the cycle (Fig. \ref{fig:SFI_Cycles}), but the Cycles 19 and 21 have by far most strong and very strong category SFI days. Note that the Cycles 19, 21 and 22 have also the strongest overall cross-correlations between SFI and Ap-index in Fig. \ref{fig:SFI_Ap_corr}. The longest Cycles 20 and 23 are very similar in the overall distribution and they have about same amount strong SFI days. Their difference is that Cycle 20 has much more moderate SFI days and much less very strong SFI days compared to the cycle 23. As expected the Cycle 24 is the weakest cycle in every respect, although it has 30 very strong SFI days comparable to the 39 very strong days of the Cycle 20. The Cycle 18 is somewhat a mystery, because it has so few moderate SFI days, quite a lot very strong SFI days, but less total number of SFI days than the Cycle 24 (SFI days for Cycles 18 and 24 are 1660 and 1940, respectively). There is, instead, a (quasi)annual variation in the total number of SFI days. We believe that some weak SFI days were missed in the Cycle 18 flare investigation. The SFI also obeys the so-called Gnevyshev-Ohl rule for the even-odd cycle pairs, i.e. that even cycle is weaker than the following odd cycle. This is true in SFI for cycle pairs 18-19 and 20-21 but not anymore for the cycle pair 22-23. It is, however, known that there are violations and some controversy about the G-O rule \citep{Tlatov_2013, Zolotova_2015}.

The Cycle 20 is interesting. It has second largest amount of SFI days (3036), and only Cycle 21 has more, i.e 3124 SFI days. It is, however, interesting that there are only 39 very strong days in the Cycle 20. We should remember, that the Cycle 20 was the era for the Apollo missions. We were lucky that the Apollo missions were carried out during the maximum and descending phase of the Cycle 20, which was much less active than the preceding grand maximum Cycle 19 and the succeeding Cycle 21. None of the Apollo flights encountered solar flares with powerful radiation or energetic particles. The only flight with quite strong H$\alpha$ flares was the Apollo mission 12, other flights were carried out during the periods of only weak flare events, except mission 13, which had some moderate activity days. According to our calculations, the preparatory flights 8\,--\,10 were done during the GG period of the Cycle 20. However, because extreme events are unpredictable, one of the most energetic and fastest solar flare events erupted on August 4th 1972, but again luckily between the last flights Apollo 16 and 17. On the other hand, minimum of the cosmic-rays was not as deep for the Cycle 20 as for the the Cycles 19 and 21 and the recovery phase started quite early in the descending phase of the this cycle allowing more cosmic-rays enter to the vicinity of the Earth during Apollo missions \citep{Takalo_2022_1}.

\section{Conclusions}

We have shown that the solar flare index shows a distinct Gnevyshev gap for the Solar Cycles 18\,--\,24. This is especially deep for the even cycles. This gap is very clear already in its PC1, which usually presents the most relevant features of the time series in question. It seems that the gap is distinctive for strong (15$<$SFI$\leq$25) and very strong (SFI$>$25) category SFI days for the even cycles. For the odd cycles the gap is seen already in the moderate (5$<$SFI$\leq$15) category SFI days, but is not as clear as in the even cycles for the strong and very strong category SFI days.

\begin{figure} 
 \centerline{\includegraphics[width=1.0\textwidth,clip=]{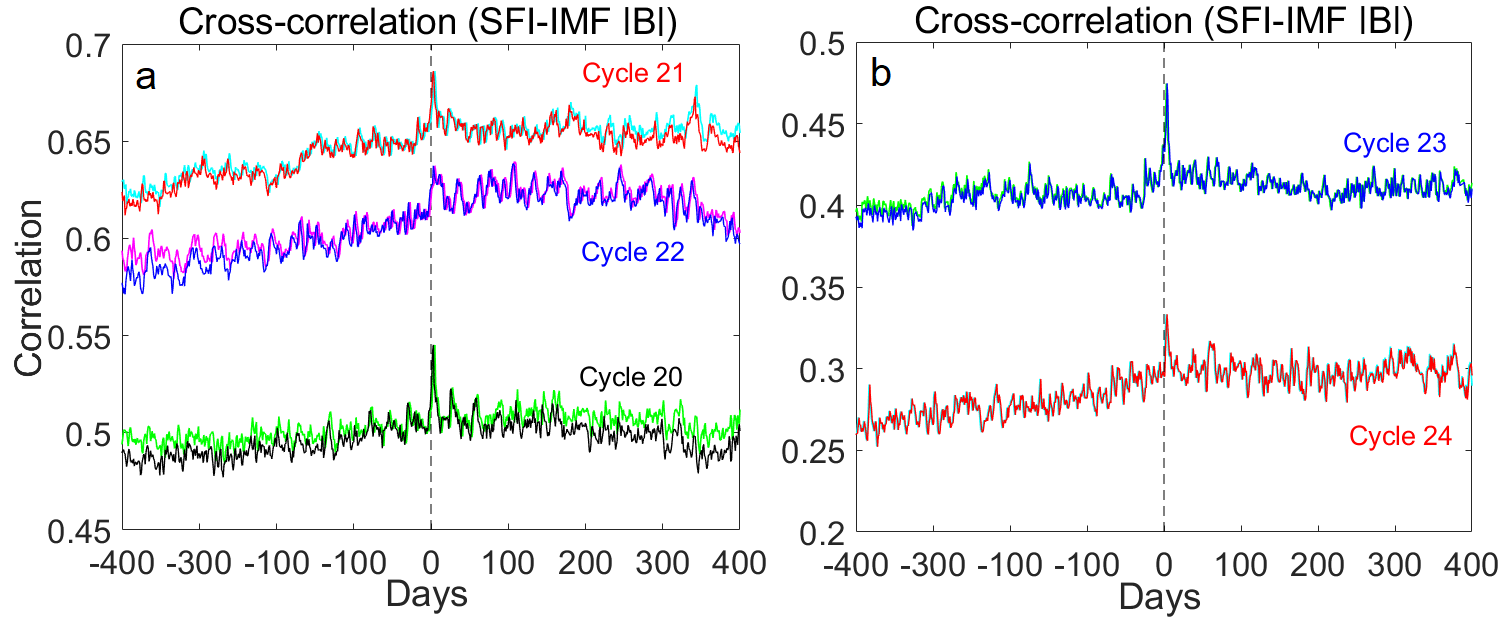}}
 \caption{The cross-correlations between SFI and IMF $\left|B\right|$-component a) for the Cycles 20\,--\,22, and b) for the Cycles 22 and 23.}
 \label{fig:SFI_IMFB_corr}
 \end{figure}

We also show that the gap is seen in the IMF $Bv^{2}$-component at the distance of the Earth and in geomagnetic Ap-index about half a year later as a deep decline. The immediate influence of the flares is seen after two to three days as a huge peak in the cross-correlation of the SFI and Ap-index. We confirm the response of the Ap-index using superposed-epoch analysis. Furthermore, the peak exists also in the cross-correlation function of the SFI and IMF $\left|B\right|$-component. Figure \ref{fig:SFI_IMFB_corr} shows the cross-correlations between SFI and IMF $\left|B\right|$ for the Solar Cycles 20\,--\,24. There are clear peaks at two to three days (SFI preceding IMF) for all cycles, except Cycle 22, which has only a small enhancement. The peaks are smaller than in the cross-correlation between SFI and Ap (see Fig. \ref{fig:SFI_Ap_corr}). It is evident that the correlation levels are in the same order as the activity levels of the SFI cycles. The colors and the methods used here are similar to those earlier in the Fig. \ref{fig:SFI_Ap_corr}. Note that the two calculated cross-correlations differ somewhat further away from the zero point, but are exactly similar at the region of the cross-correlation peak.

\begin{figure} 
 \centerline{\includegraphics[width=1.05\textwidth,clip=]{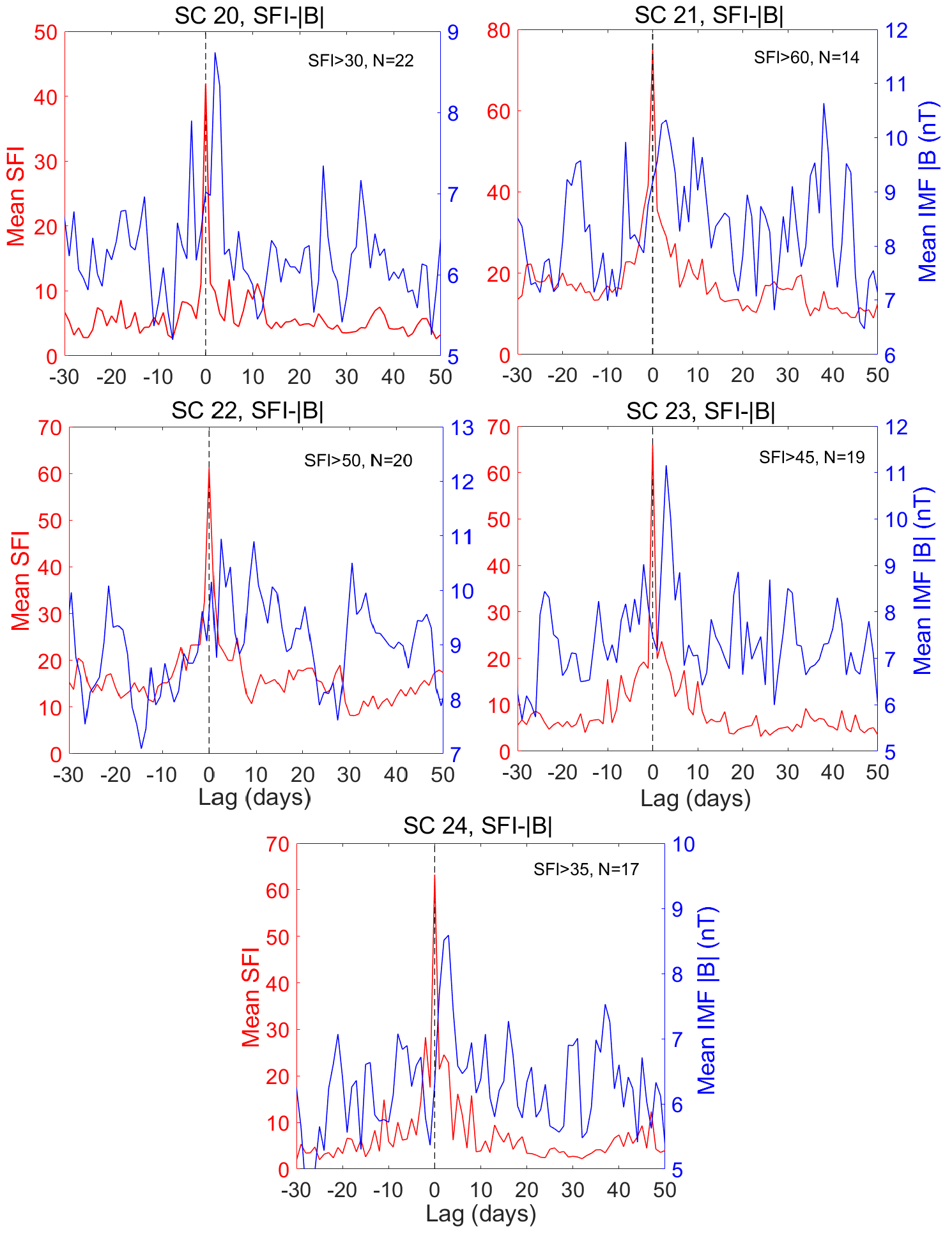}}
 \caption{Superposed-epoch analyses of SFI\,--\,IMF $\left|B\right|$-component mutual dependence for the Solar Cycles 20\,--\,24.}
 \label{fig:SFI_IMFB_epoch_analyses}
 \end{figure}

We also do similar superposed-epoch analysis for the SFI\,--\,IMF $\left|B\right|$ mutual dependence in order to confirm the aforementioned result. Figure \ref{fig:SFI_IMFB_epoch_analyses} shows these analyses for the Solar Cycles 20\,--\,24. Note that the response of the IMF $\left|B\right|$ after two to three days is very clear, except for the Cycles 20 and 21. This is because of the high overall level of the IMF $\left|B\right|$ in these active cycles. The inserted text tells again the SFI triggering level and the number of superposed epochs in each cycle.
 
The odd Cycles 19 and 21 are most active as measured with the average daily strength of SFI, i.e. 8.63 and 9.29, respectively. From the even cycles, the Cycle 22 has the highest activity in numbers of SFI days and also in the average daily value, 2917 SFI days and an average SFI 6.99. The Cycle 18 has the same amount of very strong SFI days than Cycle 22, but interestingly least nonzero SFI days from all the Cycles 18\,--\,24. We believe that something has happened in the recording of the smallest H$\alpha$ flares between Cycles 18 and 19, and there should be more nonzero SFI days in the Cycle 18. We have not analyzed the Cycle 17 earlier in this study, because recording started 1937 and Cycle 17 is incomplete. However, it seems that Cycle 17 has relatively the same amount of nonzero SFI days than Cycle 18, although it is more active in flares than Cycle 18, i.e the average SFI days are 8.37 and 5.36 for incomplete Cycle 17 and Cycle 18, respectively. Notice that in this respect Cycle 17 is almost as active than Cycles 19 and 21 \citep{VelascoHerrera_2022}. This is in line with our aforementioned suspicions. The Cycles 20, 23 and 24 are, however, the weakest flare Cycles in this study and their average SFI days are 3.30, 3.17 and 2.30, respectively.

In a detailed daily study it seemed that the GGs are more or less simultaneous in all photospheric and
 chromospheric indices for the even cycles. Only corona index (CI) has its GG later and somewhat shallower 
than for the other solar parameters. For the odd cycles the GG in SFI is somewhat earlier and less deep than 
the GG for the even cycles. For the odd cycles GG of the other indices, except PA, lag the SFI about one solar rotation period. On the other hand, the GG in PA for the odd cycles starts much later, i.e. similarly to the GG for the even cycles about 1380 days after preceding minimum of the cycle.

The GG is in solar indices, especially in the SFI, so clear that it must have influence on the space-weather as suggested earlier by \cite{Storini_2003}. If we can predict the forthcoming length of the solar cycle it is possible to presume the GG related less active time interval in the solar-terrestrial interaction.

\newpage

%% Table
%
% \begin{table}
% \caption{}%\label{tbl:?}
% \begin{tabular}{}     
% \hline
% \multicolumn{2}{c}{<>}
% <data>
% \hline
% \end{tabular}
% \end{table}

%%%%%%%%%%%%%%%%%%%%%%%%%%%%%%%%%%%%%%%%%%%%%%%%%%%%%%%%%%%%%%%%%%%%%%%%%%%
%% Appendix
%
% \appendix   

%%%%%%%%%%%%%%%%%%%%%%%%%%%%%%%%%%%%%%%%%%%%%%%%%%%%%%%%%%%%%%%%%%%%%%%%%%%
%%Acknowledgements
%
 \begin{acks}
The Solar Flare index data were obtained from https://dataverse. \newline harvard.edu/dataset.xhtml?persistentId=doi:10.7910/DVN/U5GR3D. Part of the Flare Index Data used were calculated by T.Atac and A.Ozguc from Bogazici University Kandilli Observatory, Istanbul, Turkey. The dates of cycle minima were obtained from from the National Geophysical Data Center, Boulder, Colorado, USA (https://www.ngdc.noaa.gov/stp/space-weather/solar-data/solar-indices/sunspot-numbers/cycle-data/table\_cycle-dates\_maximum-\newline minimum.txt). The newly reconstructed corona indices were obtained from www.ngdc.noaa.gov \newline/stp/solar/corona.html. The corona index for Solar Cycle 24 was downloaded from www.suh.
sk/obs/vysl/MCI.htm. The SSN2 data has been downloaded from www.sidc.be/silso/datafiles, and the solar radio
flux data from lasp.colorado.edu/lisird/data/penticton\_radio\_flux/. \newline Definitive values of Ap are provided by GeoForschungs Zentrum (GFZ) Potsdam. The OMNI2 data are downloaded from https://spdf.gsfc.nasa.gov/pub/data/omni/low\_res\_omni/. Daily plage area data are from http://cdsarc.u-strasbg.fr/ftp/J/A+A/639/A88/comp\_d.dat.
 \end{acks}

%% Available additional data environments:
%% required: authorcontribution, fundinginformation, dataavailability
%% optional: materialsavailability, codeavailability
% \begin{authorcontribution}
%
% \end{authorcontribution}
%
% \begin{fundinginformation}
%
% \end{fundinginformation}
%
% \begin{dataavailability}
%
% \end{dataavailability}
%
% \begin{ethics}
% \begin{conflict}
%
% \end{conflict}
% \end{ethics}

%%% %%%%%%%%%%%%%%%%%%%%%%%%%%%%%%%%%%%%%%%%%%%%%%%%%%%%%%%%%%%
%% Bibliography
%
% Using BibTeX
%
\bibliographystyle{spr-mp-sola}
\bibliography{references_JT_SolPhys}  
%
% Without BibTeX 
% \begin{thebibliography}{}
% \bibitem[\protect\citeauthoryear{Author}{Year}]{key}
%   <bibliographical entry>
%
% \bibitem[\protect\citeauthoryear{}{}]{}
%   
%  
% \end{thebibliography}

\end{article} 
\end{document}